\def\year{2020}\relax
\documentclass[letterpaper]{article} 
\usepackage{aaai20}  
\usepackage{times}  
\usepackage{helvet}  
\usepackage{courier}  
\usepackage{url}  
\usepackage{graphicx}  
\usepackage{graphicx}
\usepackage{makecell}
\usepackage{multirow}
\usepackage{xcolor}
\usepackage{textcase}
\usepackage[caption = false]{subfig}

\begin{document}
\title{``Trust me, I have a Ph.D.'': A Propensity Score Analysis on the Halo Effect of Disclosing One's Offline Social Status in Online Communities}

\author{Kunwoo Park$^{\dag}$,~Haewoon Kwak$^{\ddag}$,~Hyunho Song$^{\S,\P}$,~Meeyoung Cha$^{\P,\S}$\\
$^{\dag}$University of California, Los Angeles, $^{\ddag}$Qatar Computing Research Institute,\\$^{\S}$Korea Advanced Institute of Science and Technology, $^{\P}$Institute for Basic Science\\
kunwpark@ucla.edu, haewoon@acm.org, hyun78@kaist.ac.kr, mcha@ibs.re.kr
}

\maketitle

\begin{abstract}
Online communities adopt various reputation schemes to measure content quality. This study analyzes the effect of a new reputation scheme that exposes one's offline social status, such as an education degree, within an online community. We study two Reddit communities that adopted this scheme, whereby posts include tags identifying education status referred to as \textit{flairs}, and we examine how the ``transferred'' social status affects the interactions among the users. We computed propensity scores to test whether flairs give ad-hoc authority to the adopters while minimizing the effects of confounding variables such as topics of content. The results show that exposing academic degrees is likely to lead to higher audience votes as well as larger discussion size, compared to the users without the disclosed identities, in a community that covers peer-reviewed scientific articles. In another community with a focus on casual science topics, exposing mere academic degrees did not obtain such benefits. Still, the users with the highest degree (e.g., Ph.D. or M.D.) were likely to receive more feedback from the audience. These findings suggest that reputation schemes that link the offline and online worlds could induce halo effects on feedback behaviors differently depending upon the community culture. We discuss the implications of this research for the design of future reputation mechanisms.
\end{abstract}

\section{Introduction}

\noindent
Online communities strive to encourage high-quality content from their members; however, judging the quality of myriads of content has been a great challenge~\cite{brandtzaeg2008user}. A popular solution to this problem is reputation tracking whereby community members evaluate the quality of the content generated by other members by giving votes, and the aggregated votes constitute each member's \textit{reputation}. Studies have found that crowdsourced votes from peers positively correlate to content quality~\cite{stoddard2015popularity}. However, this reputation mechanism has a critical limitation. That limitation is known as the cold start problem~\cite{lampe2004slash} whereby there is no prior record for newly joining members or freshly uploaded content. Social influence and pre-existing records can also bias individuals' behavior when judging content quality~\cite{muchnik2013social,aral2014problem,berry2017discussion}.

\begin{figure}[t]
\centering
\includegraphics[width=\linewidth]{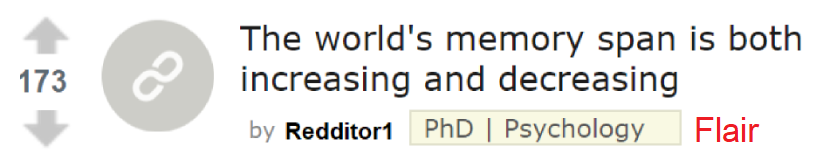}
\caption{An example of flair information}
\label{fig:flair}
\end{figure}

An alternative to building a reputation online from scratch is to ``bring'' an established \textit{social status} from the real world (e.g., academic degree or job affiliation). For example, on the question-and-answer site Quora, people can choose to reveal their domain expertise by listing job affiliations. Such ``transfer'' of offline to online status gives its members an ad hoc social status. This transfer provides additional information about the writer's knowledge, which helps the other members to better judge the credibility of the content. This mechanism may be promising because domain experts can promote their content better than others and, at the same time, overcome the cold start problem. However, this belief has yet to be tested in a data-driven manner. \textit{How does a community respond to the act of revealing offline social status such as academic degrees?}

This research brings attention to two online communities on Reddit that utilize information about members' academic degrees and domain of expertise to promote quality content:~  \emph{r/Science}\footnote{https://www.reddit.com/r/science} and \emph{r/EverythingScience}\footnote{https://www.reddit.com/r/everythingscience}.

The members in these communities use tags on posts and comments, called ``{flairs},'' as demonstrated in Figure~\ref{fig:flair}. To acquire a flair, the members must send proof of certification, such as a diploma, to the community moderators for verification. According to the board announcement, the purpose of the flair mechanism was ``to enable the general public to distinguish between an educated opinion and a random comment without a background related to the topic.''\footnote{Do you have a college degree or higher in science? Get flair indicating your expertise in /r/science!~{http://tiny.cc/2wqlqy}} 
As of March 2020, the number of subscribers in these two communities was 23.6 M and 218 K, respectively. Gathering data from these two large communities enables us to investigate how the reputation mechanism that links offline and online worlds works in the wild. We investigate the transfer of offline social status to the online world with a data-driven approach and compare the findings between the two communities.

The analysis of million-scale data over several years can identify how flairs affect the feedback behavior of community members. However, confounding factors, such as topics of content and user reputation, present challenges in measuring the effects of exposing academic status. Therefore, we conducted a propensity score analysis to infer the effect of flairs while controlling for the influences of the confounding variables on the community feedback.
We found that r/Science members who expose their academic degrees receive a more substantial amount of feedback from the other members against those who have used the community over a similar period and posted content of the same quality without any flairs. Among the users with flairs, the average effect size of exposing the highest academic degrees (such as a Ph.D. or M.D.) was three times greater than that of showing lower academic degrees such as an M.S. or indicating graduate student status. In r/EverythingScience, which covers more casual topics on science, the benefits of revealing the offline social status disappeared but remained only for the highest degree group.

Our research sheds light on the design choices for reputation mechanisms for future online communities. Exposing one's offline social status may induce cognitive bias on the perceived level of content quality, such as the halo effect~\cite{kahneman2011thinking}, which in turn can draw a disproportionate amount of community feedback to only certain content produced by individuals with high status. While this skewed attention might have been intended, it could make it difficult for online communities to promote high-quality content on the merit of the content and therefore calls for more careful designs in introducing offline status online. We hope that our findings will benefit the practitioners and designers of online reputation systems. 

\section{Related Works}
\subsection{Community Feedback as a Key for Participation}

Motivating users to actively participate and contribute high-quality content is critical for a thriving online community~\cite{malinen2015understanding}. Among the various factors that facilitate member participation, usage motivation~\cite{arguello2006talk}, personality traits~\cite{nov2013exploring}, and social networks have been found to play a role. For instance, social support boosts future engagement, as demonstrated in a study in which people committed to long-term health behaviors when they form connections~\cite{park2016persistent}. However, social networks can discourage user participation when members receive negative feedback from the community, as shown in a study related to toxic behavior in online games~\cite{shores2014identification}. Negative feedback can be fostered throughout a community, as negatively evaluated users are likely to rate others more negatively~\cite{cheng2014community,cheng2015antisocial}.
These studies suggest that community feedback and particularly a positive perspective are crucial for user participation in online communities, which is broadly the topic of this research.

\subsection{Level of Anonymity}

Significant research has investigated the anonymity level of individuals within a community. One extreme is full anonymity where one member cannot be distinguished from another by any identifier. Complete anonymity is detrimental for the credibility of a system~\cite{rains2007impact}, and it provokes a negative culture involving toxic behaviors~\cite{kilner2005anonymity,suler2004online,kwak2015exploring}. 
Some studies find that full anonymity can promote open conversations in public discourse~\cite{bernstein20114chan}. The other extreme is complete openness in which one's offline identity is exposed in the online world. Complete transparency may increase the trust and accountability of the system~\cite{kusumasondjaja2012credibility} and promote a polite communication culture~\cite{millen2003identity}.
A study investigating Amazon reviews reported that users more positively rate reviews containing identity-related information~\cite{forman2008examining}. Another study showed that fake reviews could be better detected when the author's identity information is disclosed~\cite{munzel2016assisting}. However, full openness may restrain individuals from freely sharing their views and acting naturally, as several studies have reported that privacy concerns might hinder member participation~\cite{frost2014anonymity,liao2012your}.

Most online communities advocate pseudonymity, representing a middle ground between complete anonymity and full openness, in which members create pseudonyms to build their identity. However, pseudonymity has a downside because members may develop more than one character by creating multiple user accounts, which are called \textit{sockpuppets}~\cite{kumar2017army}. Not all sockpuppets are harmful as multiple accounts are sometimes useful for people seeking social support~\cite{andalibi2016understanding,de2014mental} by allowing them to build temporary identities~\cite{leavitt2015throwaway}. Nonetheless, malicious attackers can exploit such multiple accounts with fake identities to harm other community members~\cite{wang2013you}.

\subsection{Reputation and Social Status in Online Communities}

Reputation systems promote quality content in online communities. For example, the StackOverflow website aggregates votes on the historical answers of each member as a measure of reputation~\cite{bosu2013building}. Similarly, Reddit employs a reputation system called ``Karma'' based on the voted scores of its members' historical activities. Communication studies report that online reputation creates social status---i.e., `an actor's relative standing in a group by prestige, honor, or deference'~\cite{sauder2012status}---within an online community, which further drives active participation and altruistic behaviors by motivating members to achieve higher status~\cite{lampel2007role,bateman2011research}. Online reputation systems can also help members to more easily identify experts on a given topic. Studies concerning Yahoo! Answers~\cite{shah2010evaluating} and StackOverflow~\cite{movshovitz2013analysis} have shown that online reputation scores are directly predictive of answer quality in question answering communities. 

The main challenge in implementing any online reputation mechanism is the cold start problem. One method to resolve this weakness is to transfer the offline social status, which has a hierarchy that correlates with expertise, such as academic profiles and job affiliations. MathOverflow.net uses real names instead of pseudonyms, which allows the members to match each other's offline identity to the online profile. A study involving 3,470 users of MathOverflow found a correlation between voting scores and offline social status~\cite{tausczik2011predicting}, suggesting that transferring offline status to online communities might be useful in promoting content with excellent quality.

Nonetheless, unexpected consequences might occur that could hurt the health of online communities. Social status can be divided into hierarchical levels. For example, a Ph.D. is higher than an MS, and a senior engineer is higher than an entry-level engineer. Introducing this relative difference to an online community provides a new social status structure among the members. Hence, community members can evaluate quality more generously when a user with high status posts content as observed in an interview-based study that found that Quora users perceive answers written by experts who have first-hand information on a topic to be more authoritative~\cite{paul2012authoritative}. If users with high-status receive more feedback than general users when they post content of equal quality, this may not be a desirable outcome, particularly in systems that address many niche areas of content. Despite its importance, to the best of our knowledge, no prior study has conducted an in-depth examination of the effect of introducing offline social status online on the feedback behaviors of online community members. This study aims to fill this gap using data-driven approaches.

\section{Problem and Data}

\subsection{Problem Definition}

We pose the following research question:

\begin{quote}
\itshape RQ. Does exposing one's offline social status to an online community lead to more feedback?
\end{quote}

To answer this question, we utilized logs from Reddit, which is a link-sharing and discussion website that has over a million sub-communities dedicated to specific subjects called \textit{subreddits}. Subreddit names are prefixed with `r/,' such as r/Sports or r/Gaming. Reddit members communicate by sharing web links and commenting on shared links. Each subreddit has a small group of moderators who oversee the shared content. The moderators decide the rules (e.g., terms of use violation) and pinned content (e.g., member notices). The moderators can also define the tags and flairs --- what members can show next to their profiles or posts. For example, in the r/loseit subreddit, where members share the common goal of losing weight, members adopt a flair that displays their weight loss progress (e.g., ``-50 lb''). Such flairs and tags have been shown to exert a positive peer effect~\cite{cunha2017warm}. 

In this study, we refer to the two Reddit communities, r/Science and r/EverythingScience, as \textsf{Sci} and \textsf{Eve}, respectively. The topics of these communities are either scientific manuscripts or news articles related to scientific research. The community members judge the quality of the shared content by voting up or down and, if they wish, they can participate in the comment threads associated with each post. The code of conduct is similar to that in other Reddit communities; however, their topics are limited to science. While a post may link to anything in \textsf{Eve} as long as its focus is on science, \textsf{Sci} posts are limited to peer-reviewed scientific articles that have been published in the last six months. Interestingly, the moderators of these subreddits have adopted a flair mechanism that lets members expose their education degrees and domains of expertise, such as `PhD $\mid$ Psychology', as shown in Figure~\ref{fig:flair}. 

The logs containing profiles of users with different degree types and the kinds of feedback their content received lend these subreddits to a natural experiment. Analyzing the flair dataset is advantageous for several reasons. First, the behavioral traces span nearly four years. Therefore, we are able to cover various types of scientific content over time, which is not feasible in randomized trials or interview studies. Second, their topics are limited to science, and hence, the effects of the content topics are better controlled. Third, the two similar-yet-different communities provide an opportunity to investigate the common or differing impacts of exposing offline status online. Notably, this paper is based on data-driven approaches and, hence, allows us to objectively measure the changes in community feedback without the risk of possible biases such as the social desirability bias~\cite{nederhof1985methods} that can arise in surveys or interview-based studies.

{
\begin{table*}[ht]
\centering
\begin{tabular}{c|crrcrcc}
\hline
Subreddit & Type & 
\makecell{Data entry\\(count)} & \makecell{Users\\(count)} & 
\makecell{Average per user\\(count)} & 
\makecell{Mean\\(score)} &
\makecell{Median\\(score)} &
\makecell{Std. Error\\(score)} \\\hline
\multirow{2}{*}{r/Science} & Posts & 193,441 & 73,233 & 2.641 & 144.04 & 1.0 & 3.302 
\\
& Comments & 2,487,480 & 543,524 & 4.576 & 9.446 & 1.0 & 0.434  \\\hline
\multirow{2}{*}{r/EverythingScience} & Posts & 55,966 & 12,724 & 4.398 & 20.842 & 3.0 & 0.059 
\\
& Comments & 130,542 & 31,237 & 4.179 & 5.757 & 2.0 & 0.091  \\
\hline
\end{tabular}
\caption{Descriptive statistics of posts and comments in r/Science and r/EverythingScience (Period: 2014/01-2017/12)}
  \label{table:data_desc}
\end{table*}
}

\subsection{Data}

The data analyzed were obtained from a well-known database that operates on the Reddit API~\cite{baumgartner2020pushshift}. We downloaded the primary action logs, posts, and comments that appeared on the two subreddits since the launch of \textsf{Eve} in January 2014 until December 2017. The dataset contains information regarding (1) the authors' information (e.g., name and flair); (2) the content information (e.g., identifier, timestamp, text content, and net score); (3) the crawl information (e.g., crawled time), etc. 

We carefully cleaned and sanitized the data. One challenge was to estimate the exact time when a user adopted a flair because the Reddit API does not provide this information. However, the crawled dataset contained an individual's flair status at the time of data collection instead of the time of posting. We downloaded multiple snapshots of the Reddit crawls and repeatedly checked each user to determine the times when she was last seen without a flair. Finally, for each user, we could identify the earliest data collection time when her post appeared with a flair, and we utilized all logs only after that time point. Our decision to exclude the data points is a conservative choice that guarantees reliable flair information at the time of posting.\footnote{None of the posts without a flair were removed from the data.} This step removed 6,224 and 2,977 posts and 86,078 and 7,104 comments from \textsf{Sci} and \textsf{Eve}, respectively. The Reddit snapshots had been crawled regularly over several years, with a median time difference of 27.2 days. Hence, the flair adoption time could be estimated with this margin of error.

Table~\ref{table:data_desc} summarizes the data statistics after the above filtering step and reveals that we have ample data regarding the posts and comments for the analysis. The score distributions were heavily skewed, and only a small proportion of the posts were popular, as has been observed in other Reddit communities~\cite{gilbert2013widespread}. The median scores of the posts and comments in \textsf{Sci} are both 1.0.\footnote{The initial score of the comments in Reddit was 1.} The median scores are slightly higher in \textsf{Eve}, i.e., 3.0 for posts and 2.0 for comments. These statistics do not diverge much across the whole period, suggesting that community feedback regarding content is comparable over time.

{
\begin{table}[ht]
\centering\small
\begin{tabular}{l|rrrrrr}
\hline
\multirow{2}{*}{Type} & \multicolumn{2}{c}{User} & \multicolumn{2}{c}{Post} & \multicolumn{2}{c}{Comment} \\
  & \multicolumn{1}{c}{\textsf{Sci}} & \multicolumn{1}{c}{\textsf{Eve}} &
  \multicolumn{1}{c}{\textsf{Sci}} & \multicolumn{1}{c}{\textsf{Eve}} & \multicolumn{1}{c}{\textsf{Sci}} & \multicolumn{1}{c}{\textsf{Eve}}\\\hline
DR (Doctoral) & 940 & 129 & 895 & 203 & 20,871 & 467 \\
MS (Master) & 657 & 69 & 689 & 893 & 5956 & 405 \\
GS (Grad Student) & 1,104 & 112 & 567 & 270 & 10,370 & 362 \\
BS (Bachelor) & 1,185 & 72 & 233 & 23 & 6,136 & 196 \\
   \hline
\end{tabular}
\caption{Distribution of degree types in \textsf{Sci} and \textsf{Eve}}
  \label{table:degree_domain}
\end{table}
}

The academic degree information was extracted from the collected flair tags via regular expression. We grouped similar degree types; for instance, all variants of doctoral degrees (e.g., Ph.D. and PharmD) and positions requiring an equivalent degree (e.g., Professor) were grouped. After iterative processes, we obtained the following four main degree types based on the natural hierarchy: DR (doctoral degree), MS (master's degree), GS (currently a graduate student), and BS (bachelor's degree). We further validated this automated process with the correct labels of the sampled users, which were provided by the moderators of those subreddits. Table~\ref{table:degree_domain} describes the final counts of 3,886 flair users, whose degree information was successfully retrieved. A user can only achieve a flair in \textsf{Eve} through the process of the \textsf{Sci} subreddit by the community design such that flair users in \textsf{Eve} are automatically a subset of \textsf{Sci} flair users. In the analysis, we exclude one flair type, i.e., the AMA (Ask Me Anything) flair, because it is given to science celebrities who draw a considerable amount of attention from the community.

This work is an observational study based on data gathered through the public Reddit API; hence, obtaining IRB approval is not necessary. The researchers did not intervene with the Reddit users nor process identifiable private information in this study.

\section{Direct Comparison Across Flairs}

\begin{figure*}
\hspace*{-19mm}
\subfloat[The r/Science subreddit (\textsf{Sci})]{\includegraphics[width=1.2\linewidth]{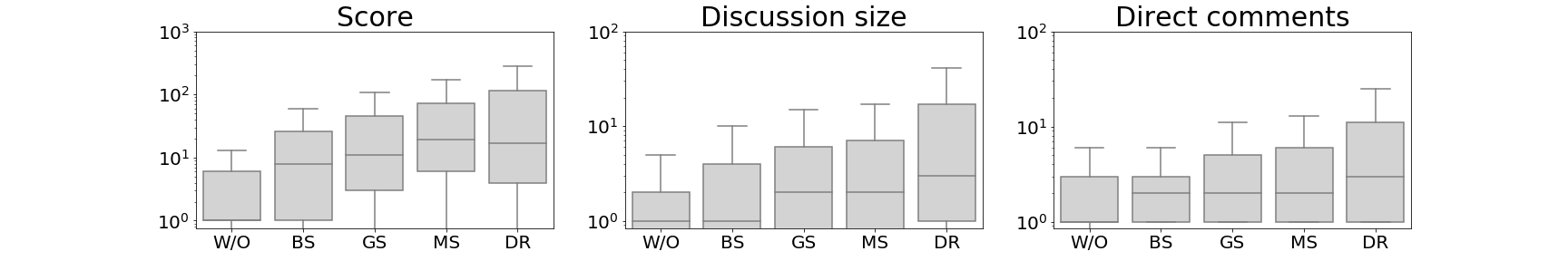}} 
\\$~$\\
$~$\\
\hspace*{-19mm}
\subfloat[The r/EverythingScience subreddit (\textsf{Eve})]{\includegraphics[width=1.2\linewidth]{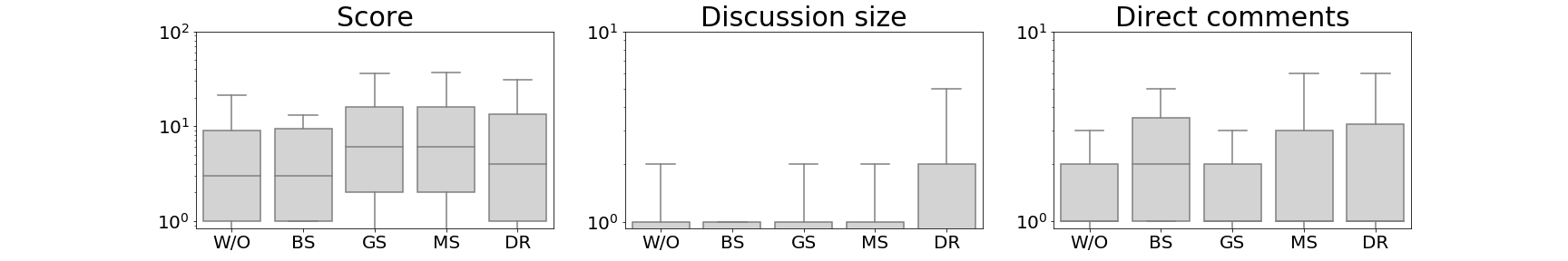}} 
\caption{Distributions of the community feedback measures across academic degree groups. ``W/O'' represents no flair.}
\label{fig:outcomes}
\end{figure*}

To answer the research question of whether flairs lead to any changes in member response, we first compared the amount of community feedback against the levels of academic degrees exposed via flairs. The following features can quantify the amount of community feedback:

\begin{itemize}
\item \textit{Score} (integer value): The net scores of the target post are calculated as \#upvotes $-$ \#downvotes. This variable captures how positively or negatively the community members evaluated the target post. While Reddit adds fuzziness to the number of upvotes and downvotes to prevent spam bots, the score, which is the difference between these votes, remains unchanged.
\item \textit{Discussion Size} (integer value): The number of comments on the target post represents the community members' participation level in the discussion.
\item \textit{Direct Comments} (integer value): 
This variable reflects the number of root comments (i.e., depth$=$1) on the discussion tree of the target post. Compared to the discussion size, this variable estimates the amount of direct feedback on the post. Posts with zero comments (45.4\% in \textsf{Sci}, 71.2\% and \textsf{Eve}) were excluded to capture a distinct pattern against the discussion size.
\end{itemize}

Figure~\ref{fig:outcomes} shows the log-scaled boxplot of the amount of community feedback that each degree group received in the two subreddits. For comparison, we show the distribution of the members without any flairs, denoted as ``W/O'' in the figure. The three feedback measures are skewed, such that they do not meet the normality assumption. Therefore, instead of using a one-way ANOVA, the Kruskal-Wallis test was employed, followed by Dunn's test as a post hoc analysis to identify the pairs of degree groups with a significant difference.

Figure~\ref{fig:outcomes}(a) suggests that exposing academic degrees through flairs is correlated with the amount of feedback received by members on their posts in \textsf{Sci}. All three measures showed a significant difference (\textit{p}$<$0.001 in each degree group combination). Compared with those without flairs, the flair posts received a larger amount of community feedback except for the bachelor's degree group (\textit{p}$<$0.001 in all comparisons). \textsf{Eve}, however, exhibits a distinct pattern in Figure~\ref{fig:outcomes}(b). Most of the users received a similar amount of feedback, and we only discovered substantial differences between the MS and GS groups. The master's degree group was likely to enjoy a higher score than those without a flair (\textit{p}$<$0.001); however, it drew a smaller discussion size (\textit{p}$<$0.001). Additionally, the posts written by the current graduate students were likely to receive a higher score (\textit{p}$<$0.001) and engage a larger audience in the discussion (\textit{p}$<$0.001) compared to those that did not expose their status. 

The above results demonstrate that a type of academic degree is likely to yield different effects. \textsf{Sci} is a clear example where the amount of community feedback correlates with the levels of academic degrees. At a glance, the group's members seem to judge the posts by high degree holders to be more attractive, as denoted by the enhanced feedback. However, users with ordinary degree holders, such as a bachelor's degree, receive even less feedback than those without any flairs.

However, the more general-science-oriented \textsf{Eve} community did not show the same pattern. We also note that there might exist confounding factors in these observations as has been reported by the previous studies on the effect of content quality~\cite{cheng2014community,keneshloo2016predicting,singer2016evidence} and user reputation~\cite{bosu2013building,shah2010evaluating,movshovitz2013analysis}. What if those with a doctoral degree received more responses not because of their flairs but because their content was significantly more interesting? These degree holders may have posted better content that discusses the latest and most significant advances in science, which the \textsf{Sci} community values. In contrast, academic degrees may not be a determinant of a successful post on the \textsf{Eve} community because anyone can engage in general posts related to science.

{
\begin{table}[ht]
\centering
\begin{tabular}{l|rrrr}
\hline
 ~~~~Community
 & \multicolumn{2}{c}{Content Quality} & \multicolumn{2}{c}{User Reputation}\\
 ~~~~~~Feedback & \multicolumn{1}{c}{\textsf{Sci}} & \multicolumn{1}{c}{\textsf{Eve}} &
  \multicolumn{1}{c}{\textsf{Sci}} & \multicolumn{1}{c}{\textsf{Eve}}\\
 \hline Score & 0.394 & 0.133 & 0.449 & 0.110 \\
Discussion Size & 0.309 & 0.113 & 0.232 & -0.075\\
Direct Comments & 0.322 & 0.140 & 0.296 & 0.041\\
\hline
\end{tabular}
\caption{The Spearman's correlation of content quality (measured by lexicon count) and user reputation (measured by aggregate score) on community feedback.} 
  \label{table:correlation_confounding}
\end{table}
}

To test the possibility of confounding effects by content quality and user reputation, we measured the (1) lexicon count and (2) estimated user karma\footnote{As the Reddit API provides only user-level karma scores at the time of API call, we estimated it as the cumulative voted scores of past posts uploaded by the same user until the time of post upload in each community.} of every post, and we investigated their correlations with the three measures on community feedback. Table~\ref{table:correlation_confounding} presents the Spearman's correlation of post quality and online reputation with our measures on community feedback. In \textsf{Sci}, the amount of community feedback exhibits a moderate and high level of correlations with content quality and online reputation, which make it possible to confound the effects of academic degrees on community feedback that was found in Figure~\ref{fig:outcomes}. \textsf{Eve} only exhibits a negligible level of correlations, and, in combination with the previous findings in Figure~\ref{fig:outcomes}, these findings show that there are no clear signals that affect the size of community feedback in the subreddit.

The results in this section showed that there is a correlation between exposing academic degree and community feedback; however, the findings also suggest that content quality and user reputation could confound the relationship. A Reddit post may receive great feedback because of the flairs or because of the confounding variables. In the following section, we apply a method to control this unwanted signal in our observational data.

\section{Estimating the Effects of Flairs}

To infer the effects of exposing one's academic degree on community feedback while controlling for the confounding influences of the covariates such as content quality features, we utilized the propensity score matching framework~\cite{rosenbaum1983central}, which is widely used in observational studies due to its ability to mitigate selection bias~\cite{guo2015propensity}. The framework consists of three steps: (1) propensity score modeling, (2) propensity score matching, and (3) estimating a treatment effect after a successful balance check. First, for each scenario that aims at testing whether exposing one of the degree types, defined in Table~\ref{table:degree_domain}, affects the amount of community feedback, a propensity model is trained to estimate the likelihood of having the treatment condition (exposing a degree type) from covariate features. Second, for each post with the treatment condition, one or more appropriate instances are matched among the control groups, which are posts without a flair, by utilizing the treatment probability estimated via the propensity model. Third, we check whether the covariate distribution of the treatment group is statistically identical to that of the control group. If it passes the balance test on the covariate distribution, the propensity score matching framework enables the estimation of the effects of the treatment condition on community feedback. For the case in which a treatment group and its matched control group have different distribution on any of covariates, the analysis framework does not allow for the estimation of the effect of a treatment condition.

The statistical analysis framework separates out the effects of the treatment conditions (i.e., exposing one's academic degree) on community feedback from the confounding influences of the covariates (e.g., content quality features), by matching appropriate instances within the control groups to each treatment unit (i.e., degree group). After this step, the covariate distribution of the treatment group should become statistically identical to that of the control group. This process approximates randomized controlled trials in which the treatment and control groups are randomly distributed with regard to covariates. Hence, the risks of confounding effects due to covariates are minimized.

\subsection{Propensity Score Modeling}

The matching process is based on the estimated propensity of each post to expose each degree type through a flair (i.e., DR, MS, GS, and BS) compared to posts without flairs as the control group. The propensity score can be modeled by any function that produces a likelihood of receiving treatment from covariates, ranging from 0 to 1. We utilized a logistic regression with Lasso regularization ($\lambda=0.001$) because this approach is known to identify essential features among a large pool of variables in the online community research~\cite{cunha2017warm,park2017achievement}. This step models the propensity scores after discarding less important covariates that can vary against experiment settings.

\subsubsection{Covariate Features}

Propensity score analysis makes the conditional independence assumption of causal inference~\cite{cunningham2018causal}, suggesting that the likelihood of having a treatment condition must be almost the same as that of being random. That is, to estimate a treatment effect accurately, one should model the propensity of receiving treatment using as many confounding variables as possible. Therefore, in addition to lexicon count and estimated user karma, which correlates with community feedback in the previous section, we considered the following variables as covariates, which quantify content quality~\cite{cheng2014community,keneshloo2016predicting,singer2016evidence} and user reputation~\cite{bosu2013building,shah2010evaluating,movshovitz2013analysis} from diverse perspectives.

\begin{itemize}
\item \textit{Catchiness} (numeric value): Catchy titles entice a larger audience and thus are likely to incur feedback regarding the target post. We measured catchiness by applying a pre-trained machine learning classifier~\cite{chakraborty2016stop} that detects clickbait news headlines to post titles. Since the model was trained based on news articles, we manually tested its adaptability to posts in \textsf{Sci} and \textsf{Eve} by sampling 50 post articles from each subreddit. The results showed a moderate agreement rate of 0.46 and 0.32 as measured by Cohen's Kappa,
suggesting that the pre-trained model can estimate the catchiness of the post titles in the subreddits with a low margin of error.

\item \textit{Readability} (numeric value): The Gunning-Fog score~\cite{gunning1952technique} estimates the years of formal education a person needs to understand the text in the first reading; thus, this score is widely used to measure online text quality (e.g., news articles~\cite{keneshloo2016predicting} and Reddit~\cite{singer2016evidence}). A higher value indicates that a given text is written with more complex lexicons and has longer sentences. 
We also considered lexicon count as a covariate because it affects the readability of each post.
We applied the Python textstat library to the post titles to measure the two variables.\footnote{https://pypi.org/project/textstat/}

\item \textit{Sentiment} (numeric value): Sentiments conveyed through post titles are known to affect the extent to which the user is likely to click the link~\cite{tatar2014survey,reis2015breaking,ferrara2015quantifying}. We utilized positive and negative sentiments as measured via VADER sentiment lexicons~\cite{hutto2014vader}.

\item \textit{Topic Distribution} (numeric value): Certain topics might attract more member attention. We measured the topic distribution of a post by applying the latent semantic indexing~\cite{papadimitriou2000latent} topic modeling method to the title with the parameter of the number of topics set to 50. The results were almost the same across several variations on a larger number of topics (i.e., 100 and 150). 

\item \textit{User Reputation} (numeric value): A user's reputation can influence the evaluation of the future content uploaded by that user~\cite{shah2010evaluating,movshovitz2013analysis}. Reddit reveals the karma score that quantifies user reputation in the user profile, which might lead to a different amount of attention. We estimated the user-level karma score at the time of post upload to be the cumulative scores of the previous posts in Reddit uploaded by the user. Additionally, a community-specific reputation was also similarly estimated by relying only on the posts within the community.
\end{itemize}

\subsection{Propensity Score Matching}

We matched the control units (i.e., posts without flairs) to each treatment unit (i.e., posts with flairs) based on the propensity score. The primary goal of matching is to obtain a balanced set, allowing for the ruling out of the effects of covariates on the outcome variables. While there are various options for matching, such as exact matching and caliper matching, we applied the $k$-nearest neighbor algorithm ($k=5$) to each treatment unit.
The similarity is measured by the \textit{Mahalanobis distance}, which is a normalized distance measure between two targets in multivariate space.

A successful matching process should yield a balanced set in terms of confounding factors. To ensure the success of this process, we measured the standardized mean difference of the propensity scores $d_{c}$ for each covariate $c$. As a rule of thumb, two groups are considered ``balanced'' if the absolute value of the standardized mean difference is below 0.1~\cite{austin2011introduction}.
We repeatedly adjusted the hyperparameter values (e.g., $\lambda$ in Lasso, and $k$ in the nearest neighbor algorithm) until we achieved balanced matching.

\subsection{Estimating the Average Treatment Effects}

We estimate the effect of a treatment condition (e.g., exposing doctoral degrees) on each outcome variable (e.g., score and other member feedback). The estimated average treatment effect (EATE) of an outcome variable $y$ was measured by the following equation:
{
\begin{equation}
\sum_{t}^{T} \sum_{m}^{M_{t}} \Big( \frac{y_{t}-y_{m}}{N_{M_{t}}} \Big) /{N_{T}}
\label{eq:eate}
\end{equation}
}

where $T$ is a set of treatment units and $M_{t}$ is a set of control units matched to treatment unit $t$. $y_{t}$ and $y_{m}$ are the outcomes measured for $t$ and $m$, respectively. $N_{T}$ and $N_{M_{t}}$ are the numbers of treatment units and matched control sets, respectively. 
As we isolate the effects of flairs from the covariates through propensity score matching, the EATE is interpreted as the number of benefits or disadvantages that are gained by disclosing academic degrees, compared to a post with the same quality that is uploaded by a user with a similar reputation yet with no flair. 
As the size of the effect may vary by treatment units (posts), the standard error of the treatment effect is also reported to carefully interpret the results.

\subsection{Matched Results}

How much advantage (or disadvantage) does a user gain upon adopting a flair compared to another user with a similar reputation and who shares content of similar quality without any flair? Table~\ref{table:causal_effects} shows the answer to this question; the average treatment effect of having a flair was computed by the \textit{Equation~(1)}. This analysis requires finding a balanced control group for each treatment group. The balance analysis revealed that the BS group did not meet this condition, leading to bias in the match analysis. Therefore, we show only the results for the DR, MS, and GS groups. The matching analysis obtained several key findings.

The first set of observations is based on the \textsf{Sci} subreddit. The individuals with the highest education level of doctoral degree were likely to gain better feedback from the community members even when they offer content of the same quality. The average treatment effects on all three variables related to community feedback were positive with a significant magnitude, and the EATE on the post score was 188.16 with a standard error of 43.46, indicating that doctorate flair could trigger more community feedback on the post. The other education levels of master's degree and graduate also show the desired effect across the three variables but to approximately one-third EATE of the former.

{
\begin{table}[t!]
\centering
\subfloat[\textsf{Sci} community]
{
\hspace*{-3mm}
\begin{tabular}{c|rrr}
\hline
Degree & \multicolumn{1}{c}{Score} & \makecell{Discussion\\Size} & \makecell{Direct\\Comments}\\\hline DR & \textbf{188.16} (43.46) & \textbf{41.82} (8.19) & \textbf{15.77} (3.5)\\
MS & 64.65 (54.62) & 13.75 (6.64) & 3.28 (3.09) \\
GS & 67.96 (41.04) & 11.19 (5.14) & 2.34 (2.21) \\
\hline
\end{tabular}
}
\\
\subfloat[\textsf{Eve} community]
{
\begin{tabular}{c|rrr}
\hline
Degree & \makecell{~~~~~~~Score} & \makecell{Discussion\\Size} & \makecell{Direct \\Comments} \\\hline
DR & \textbf{13.79} (6.02) & \textbf{2.06} (0.78) & 0.24 (0.33) \\
MS & -0.83 (2.14) & -0.23 (0.26) & \textbf{0.32} (0.22) \\
GS & 3.3 (3.44) & 0.57 (0.43) & -1.98 (1.94) \\
\hline
\end{tabular}
}
\caption{The mean effect of exposing the academic degree with the standard error in parenthesis. The largest value appears in bold text.}
  \label{table:causal_effects}
\end{table}
}

The next set of observations is on the \textsf{Eve} subreddit. We confirm a similar positive outcome in the doctoral degree group. Still, the average treatment effect was 1 to 2 orders of magnitude smaller than that observed in \textsf{Sci}, which makes the effects on direct comments negligible. The reduced magnitude is possibly due to the difference in the popularity of these two communities (notably, \textsf{Eve} has two orders of magnitude fewer subscribers.)
In the master's degree and graduate student groups, no significant effect on community feedback was observed from exposing academic degrees online. This notable difference may be due to the group's community cupture, which covers more casual science topics, unlike \textsf{Sci}, which allows only peer-reviewed articles to be uploaded. Due to such disparity, its members may no longer be biased to hold positive perceptions with regard to the doctorate group, the members of which are expected to hold expertise in their respective domains.

In summary, the propensity score matching reveals a consistent trend across the two communities: exposing a higher education degree was likely to incur more member feedback on uploaded posts; however, the mere existence of an academic degree did not guarantee the same effect. Even though the flair mechanism that was intended to invigorate community members and help identify high-quality content produced unexpected biases concerning community feedback, its results may vary against the code of conduct in each community.

\section{Discussion and Conclusion}

The reputation mechanism is commonly used in many online communities. To overcome the cold-start problem of online reputation and further promote high-quality content, some communities borrow users' offline status from the real world, as shown in the two Reddit communities studied in this work. In r/Science, exposing academic degrees through flairs corresponds to a more substantial amount of community feedback compared to the matched posts of similar quality that are uploaded by the users with almost same reputation. Moreover, the most significant effects were observed for the highest academic degrees (e.g., Ph.D.). The \textit{halo effect}~\cite{kahneman2011thinking}, which is a type of cognitive bias where one trait contributes to the overall judgment of a person, may partially explain these underlying dynamics. While promoting the high-quality content of educated users might have been intended by the community moderators, the disproportionate amount of benefits toward the education status might cause feelings deprivation in the overall population because the other users could not obtain sufficient feedback despite sharing content of the same quality. Feeling under-served and receiving steadily less feedback can further cause users to resort to lurking within the community (rather than participating) or leaving~\cite{malinen2015understanding}.
 
The studied reputation mechanism had different effects depending on community types. The \textsf{Eve} subreddit has a subtle difference in the types of information its members are allowed to share compared with \textsf{Sci}. Whereas the latter accepts only discussions regarding peer-reviewed scientific articles, the former allows casual topics. As a result, in \textsf{Eve}, academic degrees may not achieve authority except for the highest degree group, which also obtained the largest effect size in \textsf{Sci}. The varying results across the subreddits imply that showing any offline status would not always give power to the members in any community. For example, it appears unlikely that the educational status used in those scientific communities could be useful in r/Gaming. To properly supplement the online reputation mechanism, community moderators would have to select an appropriate offline status that is well-aligned with their community culture to introduce the desired effects.

While the findings and the supporting theory in social psychology provide a plausible explanation for the effects of exposing an offline social status online, the results should be carefully interpreted based on their own merits. The use of propensity score matching enables researchers to obtain a balanced distribution of each of the covariates between a treatment group and its corresponding control; however, it cannot rule out the effects of \textit{unobserved} covariates~\cite{guo2015propensity}. Had there been another variable that significantly affected the amount of community feedback, the findings of this paper would not show the true effect of exposing academic degrees online. For example, we found that excluding the variables of user reputation from the covariates even causes the direction of the effects of exposing a master's degree to be opposite to that of the reported findings. However, based on the literature, we identified the possible factors that could affect the amount of community feedback as the covariates of the analysis framework. Therefore, we believe that this study can accurately approximate the effects of exposing academic degrees online by minimizing the impact of the significant confounders.

\subsection{Changes in Post Quality After Flair Adoption}

This paper examined the research question of how other members react to posts written by flair adopters and identified a meaningful stance change upon observing a flair. What about the flair adopters? Do flair adopters also behave differently once their academic degrees are revealed to the public? Based on the relevant studies that consider social status and text writing style~\cite{sexton2000analyzing,hymes2005models}, we hypothesize that flairs render adopters aware of their social status and, consequently, lead them to change their writing style. To test this hypothesis while minimizing the risks of temporal effects, we searched for any signal of change in post quality before and after adopting the flair (i.e., between the last post without a flair and the first post with a flair). 
\begin{figure}[t!]
\centering
\includegraphics[width=.9\linewidth]{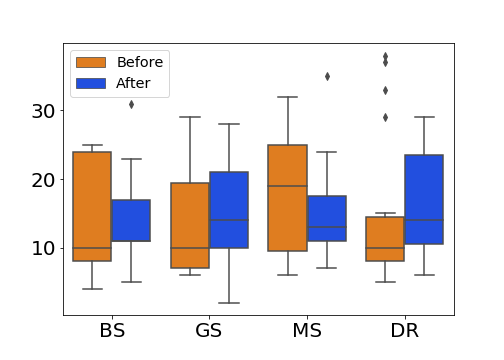}
\caption{Change of lexicon count after flair adoption. BS means users with a bachelor's degree, GS means users who are currently a graduate student, MS means users with a master's degree, and DR means users who have a Ph.D. or an equivalent degree.}
\label{fig:lexicon_change}
\end{figure}

We compared the quality features that were used as covariates in this study of the before- and after- posts of a total of 50 users in \textsf{Sci} who posted at least once before and after flair adoption.\footnote{The median post count per user is 1 in both communities.} The nonparametric Mann-Whitney U test with continuity correction was utilized for the comparison. While based on a small data sample, the writing style of the users was found to have changed once they adopted a flair. As shown in Figure~\ref{fig:lexicon_change}, the doctoral degree group began to write longer post titles once their name appeared next to `doctoral degree.' Exposing their degrees online may result in the highest social status and, in turn, cause them to feel a commitment to write post titles that are well suited to their status. We were not able to observe meaningful changes in the other groups, possibly due to the small number of users per each group.

\subsection{Limitations and Future Work}

This study examined the instantaneous authority enjoyed by members due to exposing their offline social status information to the online world. Two Reddit communities were reviewed to measure the impact of academic accomplishments: B.S. degree, M.S. degree, Ph.D. degree, and currently in graduate school. Applying propensity score measurements on this data allowed us to compare patterns across these academic achievements. However, the findings of this paper may not generalize to other types of offline social status such as occupation or affiliation. This study also bears a risk of Simpson's paradox as noted by recent studies~\cite{alipourfard2018can,lerman2018computational} in which a trend appears or disappears when aggregated. Unfortunately, we were unable to repeat the same analysis based on a more granular form of academic degree or other status type due to the small number of flair users. Future studies could test the generalizability of our findings by collecting a more extensive dataset on similar reputation mechanisms.

The chosen content quality measures bind the findings in this paper. Other methods could be applied to estimate the content quality. One could measure the coverage of news articles through the Altmetric API,\footnote{https://api.altmetric.com/} as a recent study employed this measure to identify the characteristics of popular scientific articles~\cite{maclaughlin2018predicting}. Other than propensity score analysis, synthetic controls or a difference-in-difference framework could help in the discovery of causal effects~\cite{cunningham2018causal}.
In-depth interviews could also be employed to facilitate a better understanding of the psychological impact of observing flairs on Reddit. Future research may seek to answer whether and how the ad hoc authority enjoyed by users transfers across areas (e.g., a psychology scholar posting content on physics).

Investigating \textit{who} discloses their offline status online could provide an exciting direction. Are high-status users more likely to share their offline identity? How does such disclosure affect future behaviors? Would users begin to censor their posts? While some studies report that adopting a membership badge decreases the likelihood of user churn~\cite{anderson2013steering,hamari2017badges}, we expect that introducing an offline status online could have distinct effects because it connects offline identity to online identity. Exposing a high level status might cause users to feel committed to posting more high-quality content. Members who cannot attain a high level of badges could also feel isolated and hence leave the community. Based on a sizable dataset, future studies could better explore such psychological and social effects of disclosing an offline status online. 

Another research direction is exploring how online reputation scores and offline social status \textit{interact}. In sociological theory, one's reputation enhances one's social status in the real world, and the same holds for one's online reputation~\cite{lampel2007role}. Does an online reputation have similar effects on the biases leveraged by offline social status? How does disclosed offline status affect online reputation?
Is offline social status more useful in discerning high-quality content than online reputation? A recent study found that small manipulations of Reddit post scores ultimately obtained significant changes in the end~\cite{glenski2017rating}, which suggests that previously attained scores may also induce biases toward positive evaluations. 
In the future, we plan to answer these questions to provide extensive insights into the designs of reputation mechanisms in online communities.

We are also interested in exploring the other biases that could be introduced by the studied reputation mechanism across different platforms. For instance, revealing the author's affiliations in a single-blind review policy of peer-reviewing systems has been shown to bias outcomes: reviewers perceived papers written by authors from well-known institutions to be of higher quality~\cite{tomkins2017reviewer}. Social media and microblogging platforms, such as Twitter, are used for a wide range of conversations, from personal thoughts and scientific discussions to public discourse. Given that individuals often expose their job affiliations in their profiles, we aim to determine how viewing and/or sharing offline social status affects community members' perceptions of the content they see or wish to share further in the network. 

\section*{Acknowledgement}

Majority of this work was done while K. Park was at QCRI. We thank to the moderators of r/Science who generously provided the sampled ground-truth dataset for evaluating our approach on estimating flair information. We also thank to anonymous reviewers whose comments have greatly improved this work. H. Song and M. Cha were supported by the Institute for Basic Science (IBS-R029-C2) and the Basic Science Research Program through the National Research Foundation of Korea (No. NRF-2017R1E1A1A01076400).

\bibliographystyle{aaai}
\bibliography{reference}

\end{document}